# Can extrinsic methods reveal intrinsic structure?

## Complementing IIT with QStr

Jeremiah Hendren, Matteo Grasso, Francesco Ellia, Giulio Tononi

Submitted for publication in *Qualia Structure*,
edited by N. Tsuchiya and M. Oizumi (Springer Nature)

## Abstract

Integrated Information Theory (IIT) proposes that *all quality is structure*: the quality of every experience can be characterized as a phenomenal structure and accounted for as a causal structure. The IIT method can be called *intrinsic* in that, first, it starts by characterizing the intrinsic structure of a single experience in an absolute sense (not relative to other experiences) and, second, it preserves this intrinsic perspective when accounting for experience in physical terms. By contrast, the method of the Qualia Structure paradigm (QStr) can be called *extrinsic* in that its main tool is relational characterization, used to investigate both the quality of experience and the corresponding neural or information structures.

This chapter gives conceptual clarity to IIT's intrinsic method, assesses the theoretical compatibility between QStr and IIT, and sketches concrete ways QStr can complement IIT. We argue that QStr and IIT share experience and its structure as their explanandum, while their methodological starting points and explanans are distinct yet complementary. QStr can bolster IIT's current research program by offering novel ways to approach narrow qualia (e.g., color), which are notoriously difficult to decompose through introspection. QStr can also supply IIT with new mathematical tools—for example, from category theory and metric geometry. In turn, IIT may help ground QStr's relational, extrinsic structures in absolute, intrinsic structures.

# Introduction

A trope of twentieth-century philosophy of science is that physics can tell us about relations among things but not about their "intrinsic quality" (Russell 2022 [1927]) or "intrinsic nature" (Eddington 1928). Classical arguments point to, for example, mass or charge: such constructs may help us make sense of the structure and behavior of, say, an electron, but what are mass and charge in themselves? The split of quality vs. structure has figured strongly in contemporary traditions of structural realism (Maxwell 1970; Worrall 1989; Ladyman & Ross 2007), and also in philosophy of mind. Various thinkers today entertain the idea—usually attributed to Russell—that consciousness[1] may in fact be the

[1] *Consciousness* will be treated as roughly synonymous with experience, phenomenal properties, or qualia.

"intrinsic quality" that eludes the structural analysis of the physical world (e.g., Chalmers 2003, 2015; Strawson 2006; Goff 2017; Mørch 2014; Grasso 2019; Mindt 2021). Others have resisted this move by, for example, denying that phenomenal qualities are intrinsic, private, non-structural properties at all (e.g., Dennett 1988; Frankish 2016) or outright rejecting their existence (e.g., Churchland 1986).

Integrated Information Theory (IIT) takes a different tack by collapsing the dichotomy: *all quality is structure* proposes IIT—*intrinsic* structure, that is. For IIT, *intrinsic structure* captures what a content of experience is in an absolute sense—in itself, not relative to something else. And the claim *all quality is structure* captures the central conjecture of IIT—the theory's *explanatory identity* (Albantakis et al. 2023; IIT Wiki/Identity): the quality of every experience can be characterized as an (intrinsic) phenomenal structure, which can be accounted for physically by an (intrinsic) causal structure, called a $\Phi$-structure (**Fig. 1**, steps 1–3).

As discussed in chapter 18 (Grasso et al. 2026),[2] the method of IIT can be called *intrinsic* first and foremost because it characterizes the intrinsic structure of a single experience absolutely. By contrast, most other structural approaches to consciousness (Kleiner 2024) can be called *extrinsic* in that they investigate the relationships that one experience (or content) has to others. That said, extrinsic approaches—such as the Qualia Structure paradigm (QStr, Tsuchiya et al. 2016; Tsuchiya & Saigo 2021; Maier & Tsuchiya 2026; Tsuchiya 2025)—can bolster and complement IIT's account of consciousness as intrinsic structure.

This chapter picks up where chapter 18 leaves off. It aims 1) to give conceptual clarity to IIT's intrinsic method, 2) to provide a first assessment of the theoretical compatibility between QStr and IIT, and 3) to sketch a few concrete ways in which QStr can complement IIT's current research program.

# The intrinsic method of IIT

The phrase *intrinsic structure* may appear a contradiction in terms if one's point of departure is the material world. In other words, if we assume that what exists are molecules, atoms, subatomic particles, etc., it seems nothing has intrinsic structure because every "thing" we observe can be subdivided until we arrive at fundamental entities and properties such as electrons, mass, and charge. But the notion of intrinsic structure is sensible if we start from experience itself: every moment of experience can be observed, through introspection, to be both intrinsic and structured.[3] For example, your experience reading these words is intrinsic in that it exists *for itself*, from the inside—not for something else, from the outside (IIT Wiki/Intrinsicality Axiom). And the experience can be decomposed into specific phenomenal qualities—the word you are seeing, the spatial feeling of its extension on the page, the temporal feeling of its duration as you read, etc.—each of which can be characterized as phenomenal substructures (IIT Wiki/Composition Axiom; also see Box 2: How does IIT use introspection?). This is how IIT comes by the concept of intrinsic structure: it begins not as a logical proposition about the outer world but as an introspective observation of the properties of experience as a *phenomenal structure* (**Fig. 1**, step 1).

IIT is not the first framework to analyze experience as intrinsic structure. Some aspects of this approach were already present in Kant's transcendental philosophy, in which experience is inherently structured by fundamental principles or categories (Kant 1998 [1787]). More recently, William James (1890) famously described the "stream of consciousness" as successive "pulses" of consciousness (and later

[2] Note that "chapter 18" refers to Grasso et al. (2026) throughout this preprint.
[3] These are two of IIT's five axioms—*intrinsicality* and *composition* (IIT Wiki/Axioms).

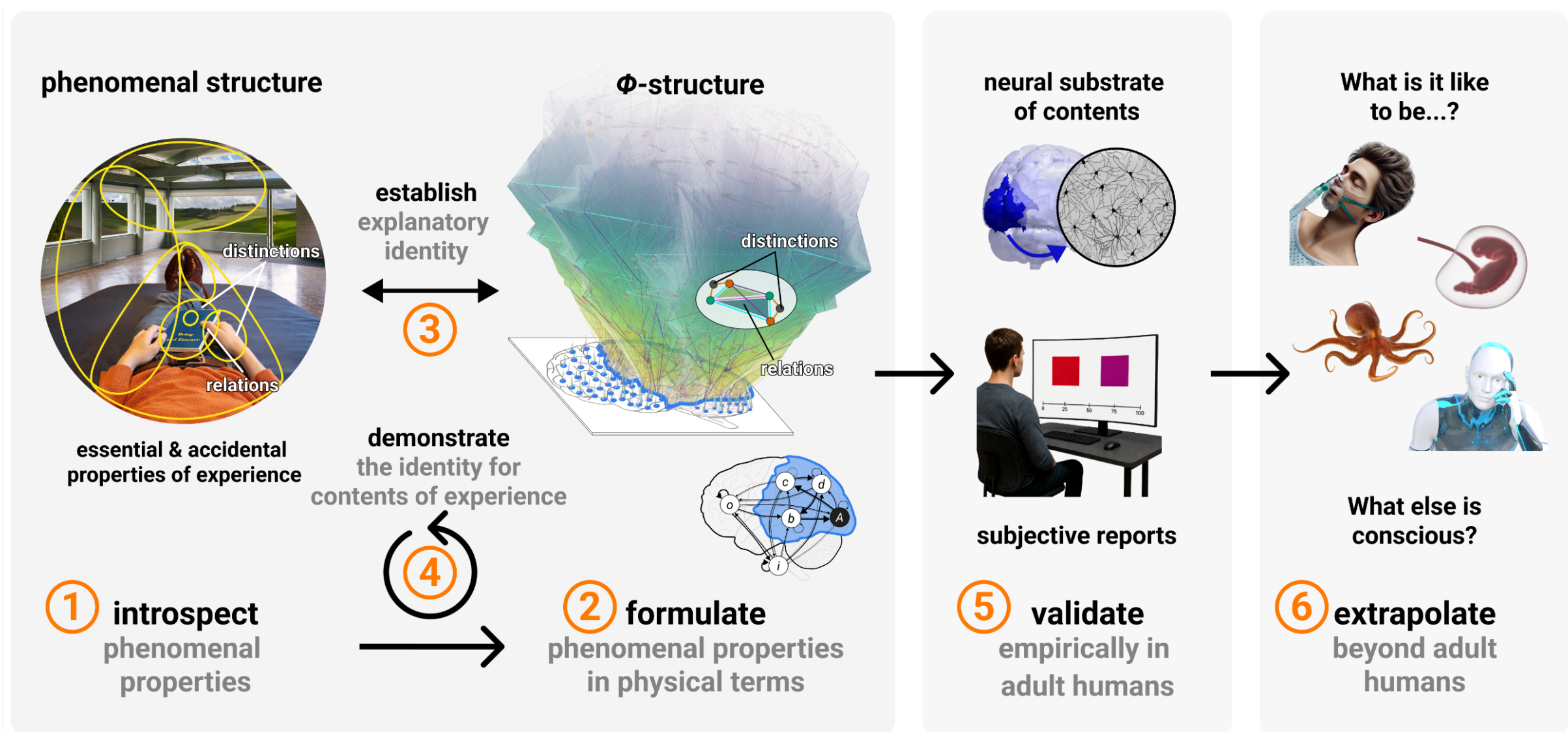

**Figure 1. The IIT Method** (adapted from IIT Wiki/Overview). **1)** IIT begins from the immediate fact that experience is present and uses introspection to identify properties that are true of every experience (*axioms*). **2)** Each essential phenomenal property is formulated as a physical (or causal) property that the substrate of consciousness must fulfill (*postulates*). This yields the notion of a $\Phi$-structure, comprising causal distinctions and relations which correspond one-to-one to phenomenal distinctions and relations, as shown schematically. **3)** IIT conjectures an *explanatory identity*: the quality of experience corresponds to the specific "shape" of the $\Phi$-structure, while the quantity of experience corresponds to its $\Phi$ value (the amount of integrated information). All essential and accidental properties of experience have their counterpart in causal properties of the $\Phi$-structure, with no additional ingredients. Experience remains primary—epistemologically and ontologically—yet the identity is *explanatory* in the sense that it offers a way to make sense of experience in objective, physical terms. **4)** IIT then uses this phenomenal–physical identity to account for contents of experience such as space, time, and objects (ch. 18). This can be thought of as repeating the previous steps (hence the "iteration" circle), but this time for accidental rather than essential phenomenal properties. Contents of experience are accounted for as $\Phi$-folds (substructures) within the greater $\Phi$-structure. **5)** IIT makes predictions that can be validated in humans who can report their experiences, validating the neural substrate of consciousness as a whole and of specific contents. **6)** IIT can guide principled inferences about consciousness in cases where reports are unavailable, such as unresponsive patients, non-human animals, or artificial systems.

"drops"), each of which could be characterized in structural terms (Natsoulas 1992, 2001).[4] Likewise, in the tradition of Phenomenology, Husserl (1982 [1913], p. 241) isolates structure as that which allows for introspection in the first place: "every mental process is so structured that there exists the essential possibility of turning one's regard to it and its really inherent components."

It would be interesting to explore further the degree to which IIT is compatible with these traditions, yet it departs from all of them in its main aim: to account for the first-person, intrinsic structure of experience in third-person, objective terms. In this regard, IIT finds affinity with a few other modern approaches. A notable example is Varela's (1996, p. 330) widely cited proposal of bridging "phenomenological accounts of the structure of experience" and "their counterparts in cognitive science" (also see Lutz & Thompson 2003). Such attempts identify experience as their explanandum and aim to characterize its intrinsic structure.[5] Nonetheless, their proposed explanans is usually defined starting from physics rather than from experience itself: it is typically a cognitive function or neural process associated with it.

[4] Though James himself did not use the term "intrinsic structure," Jamesian scholars such as Natsoulas (1992, 2001) use this term to describe James's phenomenological approach.
[5] The notion of experience as an intrinsic structure in these traditions does not always align with IIT, but we leave this aside for present purposes.

Many have argued that an explanans of this sort will never suffice to account for consciousness: the phenomenal–physical correspondence will always feel like a "brute identity"[6]—the same problem faced by the classical identity theorists of the 1950s and 60s (Goff 2017). No matter how thorough and systematic the structural correspondences are, if the explanans is constructed starting from physics, it is unlikely to "scratch our itch" of explanation; it will never fill the explanatory gap (Levine 1983) or solve (or dissolve) the hard problem (Chalmers 1995).

IIT's explanans is extrinsic and objective, like any other: it involves operational steps that we—as outside observers—perform on a substrate of units in a state (IIT Wiki/Unfolding). What differs, however, is the origin of this explanans. Following IIT's consciousness-first method (Ellia et al. 2021; Tononi & Boly 2025), we begin by identifying the essential properties of experience and then formulate each of these as a property that the substrate of that experience must satisfy in causal terms (**Fig. 1**, steps 1–2). The result is a $\Phi$-structure, which is conjectured to account for the experience in full, with no additional ingredients (**Fig. 1**, step 3). Though derived through extrinsic operational steps, a $\Phi$-structure can be thought of as an *intrinsic causal structure* since these steps deliberately "translate" the notion of an intrinsic phenomenal structure into causal terms.

To grasp the explanatory power of IIT's intrinsic method, one likely needs to see it in action for multiple aspects of experience—for example, for space, time, and objects (**Fig. 1**, step 4; as discussed in ch. 18).[7] With multiple accounts on the table, we can see how the same set of causal principles can be applied to systematically account for multiple types of experience. Each time, the notion of *intrinsic structure* provides the explanatory glue between the phenomenal and the physical. It is what allows us to reason about first-person data in third-person terms that feel genuinely explanatory. It is what makes the identity not "brute" but even "intuitive" (Hendren, in preparation; also see ch. 18 and "FAQ: What constitutes a 'good explanation' of consciousness, according to IIT? IIT Wiki/Method FAQs).

# IIT & QStr: Complementarity in theory

To demonstrate the viability of IIT's intrinsic method, we must first establish the theory's explanatory identity by going from phenomenology to physics (IIT Wiki/Identity; Ellia et al. 2021; Tononi & Boly 2025). Yet once the identity is shown to hold water, "empirical methods must then be brought in to complement introspection and reason and to extrapolate beyond their reach, in a systematic back-and-forth aiming at a good explanation of phenomenal properties in terms of physical properties" (Ellia et al. 2021, p. 10).

Among the empirical methods we might draw on, QStr may prove a fruitful way to go beyond some limitations of IIT's consciousness-first method, especially to account for narrow qualia,[8] which are notoriously difficult to decompose (see Box 2: How does IIT use introspection?). IIT calls this approach an "inference from a good explanation" (Tononi & Boly 2025) since it means starting from empirical methods (**Fig. 1**, step 5) to infer phenomenal structure (step 1). Before explaining how QStr can complement IIT in practice, we must briefly assess the compatibility of these frameworks in theory.

---

[6] This phrase "brute identity" (and variants) has been used by many thinkers, including Chalmers, Goff, and Levine. A brute identity is the assertion that two things are one and the same as a fundamental, unexplained fact, without any underlying rationale.

[7] Here, we are emphasizing how the explanatory identity is used to account for accidental properties of experience—contents such as the extendedness of visual space and the redness of red. Note, however, that the identity also accounts for the essential properties—for the presence vs. absence of experience (see Albantakis et al. 2023; IIT Wiki/Validation). The full explanatory power of the identity comes through the way a $\Phi$-structure can capture both quality and quantity of experience in objective terms.

[8] Here we adopt the distinction between "narrow" and "broad" qualia (Balduzzi & Tononi 2009; Kanai & Tsuchiya 2012; Oizumi et al. 2014; Lee-Youngzie et al. 2026). Narrow qualia are elementary contents that are not further

## A shared explanandum

To start, we believe QStr and IIT are trying to explain the same thing: experience itself (Ellia & Tsuchiya 2025, 2026)—"what it is like to be" us. This is an essential basis for complementarity since many other scientific approaches to consciousness do not take experience itself as their *explanandum* but rather the neural, functional, or behavioral correlates of experience, or the mere subjective report itself (Ellia et al. 2021; Tononi et al. 2025), which renders the explanandum itself extrinsic.

QStr and IIT also share the understanding that experiences should be characterized in structural terms. IIT formalizes this by isolating structure as one of the essential properties of experience (see ch. 18): an experience is composed of phenomenal distinctions bound by relations, and to characterize how it feels is to describe these components. QStr, too, recognizes that the feeling of a specific quale right now—say, the color red—feels the way it feels by contrast not only with other possible experiences but also with other specific qualia right now, such as the ticking of the clock.[9] For this reason, QStr methods focus on "relational characterization" (Tsuchiya 2025).

Both frameworks thus pursue what might be called a *chemistry of experience*: experience is a structure of narrow qualia bound by relations into broad qualia, much as atoms bind into molecules (see ch. 18 and Tsuchiya 2025). Whether this shared structural commitment reaches down to a shared ontology is discussed in Box 1: A shared ontology?

## Convergent methodologies

While QStr and IIT share an explanandum, their methodological starting points and *explanans* are distinct yet highly complementary.

In the IIT method, we start with introspection to define our explanatory target as precisely as possible (**Fig. 1**, step 1)—to characterize, first, the essential properties of every experience and, second, the accidental properties of specific experiences (see IIT Wiki/Overview). This first step allows us to define consciousness not as a monolith but as a phenomenal structure whose properties can be characterized in precise terms: the essential properties via introspection and reasoning on conceivability, the accidental ones via introspection of the distinctions and relations that characterize certain contents (see Box 2: How does IIT use introspection?).

IIT has focused first on characterizing the experiences of space, time, and objects because their internal structure is at least partially introspectable (ch. 18). These contents are also a sensible place to start since they are pervasive: most experiences include the feeling of visual space, of time flowing, and of objects. This strength, however, is also a source of conceptual difficulty: because these contents are so pervasive, people often fail to recognize that they are just as qualitative as the prototypical examples of

---

decomposable through introspection, such as the redness of red or the timbre of a flute. Compound contents—such as the experience of a face or a moving car—are broader qualia, with the broadest quale being the entire experience at a given moment.

[9] This echoes the notion of differentiation in IIT, which is understood as a logical consequence of the information axiom (see Box 3: Can we define what something is based on what it is not? and "FAQ: Is my experience 'specific' owing to the potential experiences I could be having?" IIT Wiki/FAQs Axioms)

qualia. The way space, time, and objects feel is no less a felt quality than the blueness of the sky or the timbre of a note, and it is just as much in need of explanation (Haun & Tononi 2019; Comolatti et al. 2025).

The QStr method also aims to characterize properties of experience but focuses on accidental ones, and employs mainly subjective reports of "relational characterizations" (Tsuchiya 2025)—for example, similarity ratings of colors (Kawakita et al. 2025; Moriguchi et al. 2025; Togashi et al. 2026) or free reports (Chuyin et al. 2024; Chuyin et al. 2025). Instead of introspecting a single content of experience, the aim is to collect subjective assessments of how qualia relate at the level of an individual subject or population.

While QStr's methods may be easier to implement, its limitation is that it captures experience only relationally, which cannot be taken as direct evidence of the identity of two experiences or contents, but only as a necessary condition for it (Kawakita et al. 2025; Robinson et al. 2025). Recognizing the complementarity of IIT and QStr, Tsuchiya et al. (2016) make the bridge explicit, proposing "a functor which relates the two domains" of qualia structures and $\Phi$-structures, and arguing that it is possible to empirically test whether such a functor exists: QStr's alignment machinery could provide constraints on the internal structure of non-introspectable contents by probing their relations to other contents, while IIT can provide an account of why that internal structure should ground those external relations.

## Complementary explanantia

Turning to the explanans in the two frameworks, IIT explicitly seeks to account for phenomenal structures through their identity with $\Phi$-structures (**Fig. 1**, step 3). IIT conceives of a $\Phi$-structure as a complete *explanans* in the sense that it allows us to account—at least in principle—for both the quantity and quality of experience in physical terms, as outlined in the previous section. In contrast, the tools and constructs of QStr do not aim to be a complete explanans in the same sense since the framework does not aim to be a theory of consciousness in itself. Its purpose is rather twofold.

First, QStr aims to provide new knowledge on the structure of experience and thus "constrain the possibility space for theories of consciousness" (Tsuchiya 2025). This is clearly achieved by obtaining qualia structures (such as those in Kawakita et al. [2025]), which give us objective insights about qualia that we cannot obtain through introspection alone.

A second, more ambitious goal is to establish commutative diagrams that describe the relations between qualia structures and information structures derived from the neural substrate (Tsuchiya 2025), thus bridging the gap between experience and its physical substrate (in this sense, it has been described as "IIT-aspirational" [Leung & Tsuchiya 2023]). Notably, as remarked in Box 1: A shared ontology?, even if thorough and systematic diagrams can be established, this does not imply any metaphysical thesis about the nature of the relations within a qualia or information structure, let alone about their potential isomorphism.

# IIT & QStr: Complementarity in practice

Based on the theoretical touchpoints above, publications are already appearing that demonstrate ways in which these frameworks complement one another (Tsuchiya et al. 2016; Tsuchiya & Saigo 2021; Phillips & Tsuchiya 2024; Maier & Tsuchiya 2026; Tsuchiya 2025; Lee-Youngzie et al. 2026). In what follows, we offer a few exploratory proposals for how QStr may support IIT's research program in the near term.

## Using QStr to bolster IIT's phenomenological methods

As outlined above, IIT insists that phenomenology is the essential starting point to bootstrap a science of consciousness (also see FAQs: IIT Method). Though necessary, this approach admittedly has limitations (Haun & Tononi 2019; Ellia et al. 2021; Tononi & Boly 2025). To say the least, introspection is constrained by our ability to attend to contents of experience, keep them in working memory, reason about them, and imagine and evaluate alternatives. IIT could no doubt profit from insights, for example, in the traditions of neurophenomenology (Lutz & Thompson 2003) and microphenomenology (Petitmengin 2006). Yet even with support from such traditions, first-person introspective methods may remain difficult to systematize (as argued in, e.g., Schwitzgebel 2008; Bayne & Spener 2010).

QStr largely sidesteps this problem altogether. The paradigm works from the elegant insight that it is easier to compare two experiential contents than to describe one. A subject simply responds to prompts in the spirit of "how similar is stimulus X to stimulus Y?" This requires little working memory, reasoning, imagination, etc.; it does not require phenomenological descriptions at all; and it can be applied to any qualia (broad or narrow). QStr still uses introspection, but it is minimal compared to IIT. As an "enriched psychophysics" (Maier & Tsuchiya 2026), QStr preserves the exact experimental control and simple behavioral outputs of psychophysics while enriching them with a greater phenomenological target—namely, the rich structure of relations among experiences. In this way, QStr avoids the burden of full first-person description, while also avoiding the poverty of single-valued psychophysical measures (see ch. 7).

On what basis can QStr sidestep introspection in this way? The paradigm has appealed to the Yoneda lemma (ch. 19; Tsuchiya & Saigo 2021): the basic idea is that any "object"—in this case a quale—can be fully characterized through the totality of its relations with other objects. For example, a qualia structure of color is obtained by asking subjects to judge different colors in contrast, never asking them to characterize a single color on its own. Nevertheless, the Yoneda lemma suggests that the final qualia structure indeed says something about, say, redness *itself*—in an absolute sense (see Box 3: Can we define what something is based on what it is not?). This way of thinking may prove especially helpful when investigating narrow qualia.

## Investigating narrow qualia

As mentioned already, a limitation of IIT's phenomenological starting point is that some types of qualia are very difficult, if not impossible, to decompose through introspection (see Box 2: How does IIT use introspection?; see also ch. 18; Haun & Tononi 2019; Ellia et al. 2021; Tononi & Boly 2025). The IIT method (**Fig. 1**) can most properly be applied to contents of experience that are at least partially introspectable, such as space, time, and objects. As broad qualia, these can be decomposed into phenomenal components much more easily than narrow qualia such as color or pain, which are often evoked to bemoan the ineffability of qualia (Dennett 1988) or to emphasize the "intrinsic residue" left out of structural accounts.[10] However, for IIT to deliver on its claim that *all quality is structure*, narrow qualia must also be shown to exist as intrinsic phenomenal structures.

---

[10] The phrase "intrinsic residue" is due to Seager (2018), but the core idea has been expressed in various ways to motivate the hard problem ("consciousness will always be a further fact relative to structural and dynamic facts," Chalmers 1996, 122) or the explanatory gap ("[there is something] crucial left out," Levine 1983, 357), in addition to numerous sub-problems such as the knowledge argument ("this mysterious residue," Jackson 1982, 136) and inverted qualia (Chalmers 1996, ch. 5).

Here, what is a limitation of IIT's introspective methods is even a strength of QStr. It is not only possible to construct a qualia structure for narrow qualia, but the method of "relational characterization" is most easily applied here. A subject can, for example, easily judge the similarity of two colors or two timbres. But the broader the qualia, the more potential confounds are introduced into QStr's methods because relational comparisons must take into account multiple dimensions. Faces, for instance, could be compared based on their spatial properties (e.g., size), the narrow qualia composing them (e.g., color), their emotional character (e.g., happy or sad), and so on.

QStr methods can thus help IIT bootstrap explanations of narrow qualia by switching directions methodologically. That is, referring back to Figure 1, we would start from qualia structures, obtained through subjective reports, and from knowledge of the corresponding neural substrate (step 5). We would then unfold the $\Phi$-fold (i.e., causal substructure) of that substrate (step 2), and use this $\Phi$-fold together with the qualia structure as a guide to infer the intrinsic phenomenal structure of these contents that cannot be introspected directly (step 1).

Let us preview what this might mean for IIT's account of color qualia—a fitting place to start since we already have qualia structures of color to work with (Kawakita et al. 2025; Moriguchi et al. 2025; Togashi et al. 2026).[11] As presented in ch. 18, IIT hypothesizes that *cliques* of color-selective substrate units could be unfolded into *kernels*—a type of $\Phi$-fold—each of which would correspond to the way each color feels (also see Tononi & Boly 2025).[12] This conjecture cannot be explored starting from phenomenology (**Fig. 1**, step 1) simply because we cannot decompose, say, redness through introspection—it just feels monolithically red. As a result, the notion of a *color kernel* is just a placeholder for now, whose intrinsic causal structure we need to discover through other creative means.[13]

Let us assume we have gained knowledge of the neural cliques supporting our experience of color. In principle, we could then unfold $\Phi$-folds from those color-specific cliques (or simplified models thereof). Presumably, the kernels would not be particularly informative at first glance. $\Phi$-folds are complex mathematical objects, and any two might be aligned in more than one way—on brightness or saturation, say, rather than hue. If the explanatory identity of IIT holds, however, fully unfolding color kernels should already specify these dimensions—a circular dimension of hue alongside circular ones for saturation and brightness—and account precisely for why red feels closer to orange than to blue.

It would be quite demanding to fully unfold color kernels in this way, but here the qualia structure of hues could be of great help. With this in hand, we could create a "commutative diagram" (Tsuchiya 2025) between the qualia structure and the corresponding $\Phi$-folds, testing whether the alignment of the $\Phi$-folds reproduces the measured phenomenal similarities among hues. We could thus isolate a specific "causal motif" (ch. 18) in these kernels—namely, the one whose alignment recovers those similarities. This motif may have some overarching properties that we might identify as definitive of a color kernel—perhaps with variants for specific hues, much like the causal motif of conceptual hierarchy (ch. 18). In sum, the QStr method would be a top-down, extrinsic, and indirect way to uncover the intrinsic structure of a content of experience—serving at once to characterize the phenomenology more fully and to supply the relational data on which any unfolded $\Phi$-structure must be validated.

---

[11] For example, Kawakita et al. (2025) collected similarity judgments of over 93 colors and used Gromov–Wasserstein optimal transport (GWOT) to align two people's color structures "without presupposing correspondences (such as 'red-to-red')," thus obtaining a qualia structure of color that seems to hold across color-neurotypical people.

[12] These cliques might be found, for example, in densely connected single- and double-opponent cells in early visual cortex mediating local cooperative and competitive interactions.

[13] We must also recognize that knowledge of the neural substrate (in this case, the hypothesized cliques) is currently limited, and that unfolding it is computationally demanding. Yet these neuroscientific and computational challenges are more tractable than the introspective one.

For the case of color, this exercise might sound trivial: we don't need to apply the complex method of QStr to obtain a model of the similarity between colors—the color spindle was available all along. Yet this is inaccurate. First, QStr methods are proving successful in assessing the accuracy of long-standing models of color space, showing, for instance, that they are largely preserved across subjects. Second, QStr methods are providing a way to detect individual variations—for instance, by modeling the color space of color-atypical subjects and their subtypes, such as red–green color blindness (see Kawakita et al. 2025).

Furthermore, the QStr method is even more fruitful when applied to other kinds of narrow qualia: cases like taste or smell may better showcase the real strength of qualia structure—not only because their internal structure is hard to introspect but because the relations *among* qualia are also much less known. For instance, some may argue it is easy to predict that most subjects would rate red and orange as similar and as both different from blue. It is, however, a non-trivial question whether most subjects would rate a dry red wine as more similar to a sweet red wine than to a dry white. One could imagine a procedure similar to the one outlined above to isolate intrinsic causal structures corresponding to, say, tastes or smells. See ch. 10 (Robinson et al. 2026) for a similar example and a discussion of how this informs comparisons across experiences.

## QStr's "information structures" as a proxy for $\Phi$-structures

Another way in which QStr can complement IIT is by connecting qualia structures with "information structures" obtained from neuronal data (Tsuchiya et al. 2016; Tsuchiya 2025; Maier & Tsuchiya 2026). QStr proposes the notion of information structure as a stopgap measure, a placeholder that can play the explanatory role of a $\Phi$-structure but that can be studied without yet performing full IIT-style unfolding. This approach has been illustrated in Oizumi et al. (2025) using *principal bundle geometry*: the idea is to use neural recordings and mathematical alignment methods to compare subjects' experiential structures indirectly, especially starting from intersubjectively stable domains such as space and color, which are easier to align because they are more "hardwired" across subjects than higher-level concepts.

Since $\Phi$-structures are quite challenging to unfold from neural data, the QStr notion of information structures could function as a proxy for the substrate-side structure relevant to *contents of experience*—the "shape" of a $\Phi$-structure—similar to how the Perturbational Complexity Index (Casali et al. 2013) serves as a practical proxy for the substrate-side structure relevant for the *level of consciousness*—the $\Phi$ value itself. Given neural recordings, one might devise a principled way to detect signatures of experiential contents—such as space or color, and eventually even pain—even if one cannot yet unfold the underlying $\Phi$-structure. This proxy is scientifically useful in itself because it could allow researchers to align neural data with qualia-structure reports and do real empirical work on experiential contents right now. Moreover, it could serve as a stepping stone toward full IIT validation later, if and when it becomes possible to unfold a neural substrate.

## QStr as a source of novel mathematical tools

QStr can complement IIT at a more formal level in two ways: by supplying a mathematical vocabulary that IIT has left implicit, and by importing tools that expand the scope of problems IIT can address. Most of this vocabulary comes from category theory (CT), the branch of mathematics that studies objects through the structure-preserving maps between them; a final tool, *optimal transport*, adds a way to compare such structures quantitatively.

*Adjunction.* Tsuchiya, Saigo, and Phillips (2023) propose the use of adjunction to characterize the relationship between the category of qualia and that of reports, capturing the systematic but lossy correspondence between what is experienced and what can be communicated about it. This construction recurs across the program: it figures in Lee-Youngzie et al. (2026) and, in this volume, in chapter 10 (Robinson et al. 2026), in connection with the "Rosetta Stone" problem of translating between physical, phenomenal, and report-level descriptions of one and the same experience.

*Sheaf and presheaf.* Lee-Youngzie et al. (2026) apply CT directly to experience. Relating to IIT's account of visual space in terms of inclusion relations (Haun & Tononi 2019), they model the visual field as a topology of mereological parts—for example, the left and right hemifields meeting at the vertical meridian. They then ask whether measures of narrow qualia over these parts can be uniquely combined into a measure of the broad quale for the whole visual field. Where this condition holds, the broad quale is fully determined by its narrow constituents, forming a sheaf; where it fails—impossible figures, binocular rivalry—locally compatible parts admit no consistent whole, forming a presheaf.

*Diversity index*. In IIT, $\Phi$ measures the quantity of experience as the sum of the integrated information of the distinctions and relations composing a $\Phi$-structure. Since $\Phi$ measures the numerosity, not diversity, of distinctions and relations, an experience of pure space may have the same $\Phi$ value as an experience characterized by many kinds of contents, and IIT has as yet no measure to tell the two apart. Kusano, Saigo, and Tsuchiya (2026) faced an analogous problem for qualia structures, importing from CT the indices *magnitude* and *spread*, which measure the effective number of distinguishable points in a single structure. Tested on similarity judgments of color and emotion words, these indices quantify the diversity—or structural "size"—of a single structure, without requiring a second structure for comparison. Applied to a $\Phi$-structure, such an index might capture the variability of an experience alongside its quantity.

*Optimal transport within and across metric spaces.* IIT 3.0 (Oizumi et al. 2014) compared causal structures using the *Wasserstein distance*, also known as the *earth mover's distance* (EMD).[14] IIT 4.0 replaced EMD with an intrinsic-difference measure for computing $\varphi_s$ ("system phi"), which does not automatically offer a principled metric for comparing $\Phi$-structures (Albantakis et al. 2023). Yet such a metric is what is needed to ask, on the substrate side, how similar $\Phi$-folds are to one another—the very question QStr answers on the phenomenal side. A candidate comes, fittingly, from QStr—in fact from the same hand that penned IIT 3.0: Oizumi and colleagues (Kawakita et al. 2025) have applied *Gromov–Wasserstein optimal transport* (GWOT)—a generalization of the same Wasserstein distance from IIT 3.0—to the study of qualia structure. GWOT seeks the best mapping between two point clouds using only the pairwise dissimilarities internal to each. Its strength is to align structures with no shared axes and no assumed labels ("unsupervised" alignment), returning both a *transport plan* and a *distance* quantifying how well the two structures fit. Because GWOT compares relational structures without a shared coordinate system, it is a natural candidate for IIT 4.0 to compare $\Phi$-structures (or $\Phi$-folds) across contents, subjects, or systems, and potentially to align qualia structures with $\Phi$-structures directly. Tsuchiya (2025) frames such cross-domain comparisons as one of QStr's most ambitious goals—that is, comparisons between "qualia and causal structures at the level of a category of categories." Since pairwise GWOT cannot capture the higher-order relations involved in experience, optimal transport is now being extended to hypergraph-like structures for direct comparison of IIT 4.0

[14] These measures were used both to compute integrated information as the distance between a system's cause–effect structure and that of its minimum information partition and, more generally, to compare cause–effect structures (then referred to as "constellations of concepts" in "concept space"; Oizumi et al. 2014).

$\Phi$-structures (Togashi et al., in preparation). If successful, this would give IIT a quantitative tool it currently lacks.

# Conclusion

This chapter has aimed to clarify IIT's intrinsic-structural approach, provide a first assessment of the theoretical compatibility between QStr and IIT, and sketch a few concrete ways in which QStr can complement IIT's current research program. We have argued that they share experience as their explanandum, and that their methodologies and explanans are distinct yet complementary. QStr can bolster IIT's use of introspection where it falls short, help reach the narrow qualia that resist decomposition, provide a proxy connecting $\Phi$-structures to neuronal data, and bring new mathematical tools into IIT's repertoire.

Though we have focused on how QStr can help IIT, it is worth noting that IIT may, in turn, prove useful to ground the relational, extrinsic structures of QStr in the absolute, intrinsic structures of IIT. Without an ontological grounding of this sort, it is difficult for QStr to offer a principled reason for why the relations it measures should hold. This complementarity can especially inform extrapolations about the perennial question of other minds (**Fig. 1**, step 6), taken up in ch. 10 (Robinson et al. 2026).

# Boxes

## Box 1: A shared ontology?

One might ask whether QStr and IIT truly share the same explanandum. If we examine QStr beyond its conceptual and epistemological tools, a stronger ontological reading of the paradigm echoes a position known as *ontic structural realism* (Tsuchiya 2025; Ladyman & Ross 2007; French 2014; Rovelli 2021). Such a position treats relations, rather than the relata themselves, as ontologically primary. This commitment is made precise by its mathematical centerpiece, the Yoneda lemma of category theory, which establishes that an object is characterized, up to isomorphism, entirely by the relations it bears to all other objects in its category (Tsuchiya & Saigo 2021). Applied to consciousness, this approach defines the phenomenal character of a quale—such as a particular shade of red—through its relations to every other experience, through its place within the global structure of qualia. IIT, by contrast, takes an experience—a phenomenal structure—to be intrinsic, existing for itself from its own perspective rather than in virtue of its relations to something else.

Do the two frameworks, then, really target the same thing? The tension may be only apparent since both frameworks take experience itself to be an intrinsic structure. To note, IIT may even furnish the grounding that is lacking in a purely relational structuralism, which faces "the Newman problem" (Newman 1928; Kleiner 2025): if qualia are individuated only by their mutual relations, the resulting structure may be trivially satisfiable and fail to capture what makes a quale exactly what it is. IIT's conjecture that "all quality is structure" addresses this directly: the phenomenal character of each quale is due to its own internal, intrinsic structure—even for seemingly simple qualia such as the redness of red, whose internal structure is impenetrable to introspection (see ch. 18). In this view, therefore, the relations QStr measures across qualia are grounded in the intrinsic structure of experience, giving QStr's quality space an intrinsic and non-arbitrary foundation and a principled answer to Newman's objection. That said, these alternative interpretations of QStr deserve greater attention in the future.

## Box 2: How does IIT use introspection?

IIT uses introspection to identify and characterize properties of experience—both essential and accidental.

To identify the essential properties (IIT's *axioms*), we isolate a candidate—say, *integration*—and then try to conceive of an experience that lacked this property. For example, if we try to conceive of an experience that were not *integrated*, we end up imagining two or more distinct experiences, each of which is indeed integrated. Since this result is contradictory, we thus confirm that the property is essential to every conceivable experience. This technique is similar to Descartes's *cogito* argumentation and to "proof by negation" in mathematics (see IIT Wiki/Axiom FAQs).

IIT's five axioms collectively allow us to conceive of any "moment" of experience as a phenomenal structure. Hence, to characterize the accidental properties of an experience means to "decompose" it into the unique types of phenomenal distinctions and relations that give the experience the specific phenomenal structure it has (as introduced in ch. 18 for space, time, and objects). This technique of decomposition is admittedly difficult to systematize. However, the guiding questions for each content are, What is the fundamental building block of this content, and how do these combine to give the specific qualitative character it has? Like in other phenomenological traditions, we aim to "bracket" all other contents of experience to isolate the one in question. For visual space, for example, we bracket the fact that space nearly always has colors, contours, or objects "painted on," focusing instead on what makes the "canvas" of space itself feel extended.

In this approach, we call broader qualia "more introspectable" than narrow qualia for the simple reason that we can partly decompose their distinctions and relations. In contrast, a narrow quale such as redness cannot be decomposed introspectively—it just feels monolithically red.

## Box 3: Can we define what something is based on what it is not?

We suspect there may be a fruitful parallel to be explored between QStr's use of the Yoneda lemma and IIT's notion of *differentiation*, which follows from its information axiom (IIT Wiki/Information). The axiom states that every experience is specific—redness, for example, simply feels the specific way it feels. Though not technically part of the axiom,[15] differentiation is a logical consequence of it: by virtue of being specific, a given experience necessarily differs from other possible experiences. IIT formulates the property of phenomenal differentiation in causal terms: a given substrate of units in a state specifies a cause–effect state, which—by virtue of being specific—differs from a repertoire of other possible cause–effect states the substrate could specify. Hence, for a given $\Phi$-structure to exist in an absolute sense, there must potentially exist a repertoire of other possible structures, regardless of whether the system has ever specified them or ever will.

[15] The axioms of IIT aim to express properties that are immediate and irrefutable based on introspection alone. See "FAQ: What is meant by the term axiom in IIT?" and "FAQ: Is my experience 'specific' owing to the potential experiences I could be having?" (IIT Wiki/Axiom FAQs)